\documentclass{aa}
\usepackage{natbib}         
\usepackage{graphicx}
\usepackage{txfonts}
\bibpunct{(}{)}{;}{a}{}{,} 
\newcommand{\ind}[1]{_{\mathrm{#1}}}
\newcommand{\diff}{\mathrm{d}}

\newcommand\Dnu{\Delta\nu}

\newcommand\Tg{\Delta\Pi_1}

\newcommand\nus{\nu\ind{s}}
\newcommand\dnurot{\delta\nu\ind{rot}}

\newcommand\numax{\nu\ind{max}}

\newcommand\epsg{\varepsilon\ind{g}}

\newcommand{\Msun}{M_\odot}
\newcommand{\Rsun}{R_\odot}

\newcommand{\momin}{I}
\newcommand{\momcin}{J}
\newcommand{\fac}{\alpha}
\newcommand{\coefI}{\beta}
\newcommand{\expoO}{\gamma}
\newcommand{\Mc}{M\ind{c}}
\newcommand{\rHBS}{r\ind{HBS}}
\newcommand{\rBCZ}{r\ind{BCZ}}
\newcommand{\Omegac}{\Omega\ind{c}}

\newcommand{\Omegaenv}{\Omega\ind{env}}

\newcommand{\alyas}{\delta\nu\ind{Kep}}
\newcommand{\Ntotal}{1500}

\newcommand\Kepler{\textsl{Kepler}}

\begin{document}

\title{Locked differential rotation in core-helium burning red giants\footnote{Table 1 is only available in electronic form at the CDS via anonymous ftp to cdsarc.u-strasbg.fr ...}}
\titlerunning{Red-clump star core rotation}
\authorrunning{B. Mosser et al.}

\author{%
 B. Mosser\inst{1},
 G. Dr\'eau\inst{1,2},
 C. Pin\c con\inst{3},
 S. Deheuvels\inst{4},
 K. Belkacem\inst{1},
 Y. Lebreton\inst{1,5},
 M-J. Goupil\inst{1},
 E. Michel\inst{1}
}

\institute{
LESIA, Observatoire de Paris, Universit\'e PSL, CNRS, Sorbonne Universit\'e, Universit\'e de Paris, 92195 Meudon, France; \texttt{benoit.mosser@observatoiredeparis.psl.eu}
\and
School of Physics, University of New South Wales, NSW 2052, Australia
\and
Universit\'e Paris-Saclay, CNRS, Institut d'Astrophysique Spatiale, 91405, Orsay, France
\and
IRAP, Universit\'e de Toulouse, CNRS, CNES, UPS, 31400 Toulouse, France
\and
Univ. Rennes, CNRS, IPR (Institut de Physique de Rennes) - UMR 6251, F-35000 Rennes, France
}


\abstract
{Oscillation modes of a mixed character are able to probe the inner region of evolved low-mass stars and offer access to a range of information, in particular, the mean core rotation. Ensemble asteroseismology observations are then able to provide clear views on the transfer of angular momentum when stars evolve as red giants.}
{Previous catalogs of core rotation rates in evolved low-mass stars have focussed on hydrogen-shell burning stars. Our aim is to complete the compilation of rotation measurements toward more evolved stages, with a detailed analysis of the mean core rotation in core-helium burning giants.}
{The asymptotic expansion for dipole mixed modes allows us to fit oscillation spectra of red clump stars and derive their core rotation rates. We used a range of prior seismic analyses,  complete with new data, to get statistically significant results.}
{We measured the mean core rotation rates for more than $\Ntotal$ red clump stars. We find that the evolution of the core rotation rate in core-helium-burning stars scales with the inverse square of the stellar radius, with a small dependence on mass.
}
{Assuming the conservation of the global angular momentum, a simple model allows us to infer that the mean core rotation and envelope rotation are necessarily coupled. The coupling mechanism ensures that the differential rotation in core-helium-burning red giants is locked.
}


\maketitle
\section{Introduction}

Solar-like oscillations in red giants have been extensively studied since their first unambiguous detection with the CoRoT mission \citep{2009Natur.459..398D}. The NASA mission \Kepler\ then provided longer time series and the yields have benefitted from enhanced methods for analysing the oscillation pattern and finer frequency resolution \citep[e.g.,][]{2009A&A...506...33M,2010ApJ...713L.176B,2010ApJ...723.1607H}.
Contrary to the Sun and low-mass main sequence (MS) stars, pressure waves excited in the upper stellar envelope of red giants can efficiently couple with gravity waves, whose propagation is restricted to the inner region delimited by the radiative core. Therefore, non-radial modes show a specific pattern of mixed modes, as theoretically predicted and depicted much before the large seismic surveys provided by space-borne missions \citep{1979PASJ...31...87S,1989nos..book.....U}. Taking advantage of these properties, dipole mixed modes have opened a new window in asteroseismology, due to their ability to probe the stellar core of red giants \citep{2011Sci...332..205B}, unveil the evolutionary stage of evolved stars \citep{2011Natur.471..608B}, measure the mean core rotation \citep{2012Natur.481...55B}, and/or detect strong magnetic fields in the red giant cores \citep{li22}.

In practice, the understanding of the mixed-mode pattern has proven to be very useful for a thorough analysis of solar-like oscillations in red giants. In the last ten years, the dipole mixed mode pattern of many red giants has been analyzed for a wealth of purposes, using the asymptotic expansion and theoretical developments \citep[e.g.,][]{2013A&A...549A..75G,2014A&A...572A..11G}.
Altogether, all the stars for which individual mixed modes were identified and fitted in previous work now form a rich, valuable data set of more than 600 stars on the red giant branch (RGB) and 700 in the red clump (RC). \cite{2018A&A...616A..24G} already presented the analysis of the core rotation on the RGB. Here, we focus on core-helium burning stars, whose importance in terms of  probing stellar densities, kinematics, and chemical abundances  in the Galaxy has been illustrated by \cite{2016ARA&A..54...95G} and references therein.

An extensive analysis of the core rotation of RC stars has not yet been made, with current information limited to the first catalog by \cite{2012A&A...548A..10M}, with about 310 red giant stars, including 225 located in the red clump. Such a data set is too small to enable a thorough study of the dependence on stellar evolution or on such parameters as the stellar mass. In this work, we use the consistent description of all seismic parameters provided by the asymptotic expansion \citep{2018A&A...618A.109M} to decipher the mixed-mode oscillation pattern of core-helium burning star to infer their mean core rotation.

In Section \ref{context}, we present the scientific context and the data used in this work. A specific seismic analysis for measuring rotation rates, based on stretched frequency spectra, is given in Section \ref{analysis}. Our results related to the mean core rotation in core-helium burning stars are discussed in Section \ref{discussion}. Section \ref{conclusion} is devoted to our conclusions.

\section{Data\label{context}}


\begin{figure}
\includegraphics[width=8.8cm]{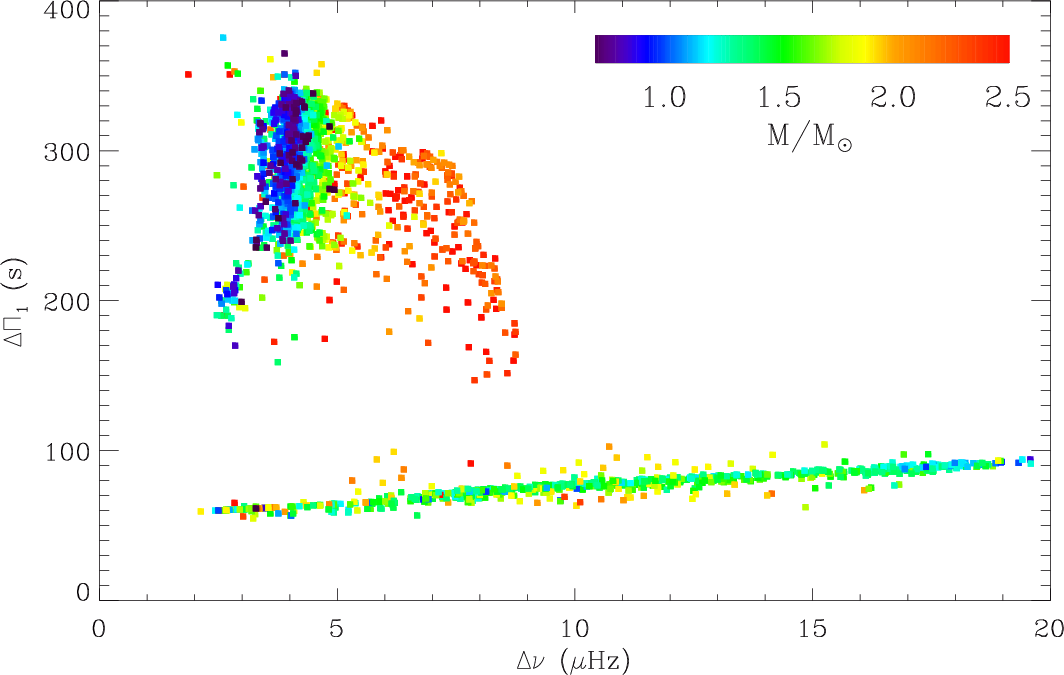}
\caption{$\Dnu$ -- $\Tg$ diagram of the core-helium burning stars used in this work. The seismic estimate of the stellar mass is color coded. The low branch corresponds to stars ascending the RGB. Most of the outliers below the global trend of the RGB were identified as stars with a mass transfer by \cite{2022A&A...659A.106D}. Typical uncertainties are below 0.1\,$\mu$Hz for $\Dnu$ and below 4\,s (1\,s) for $\Tg$ for, respectively, RC (RGB) stars.
}\label{fig-dnuDPi}
\end{figure}

Rotational splittings in RC stars can be quite small \citep{2012A&A...548A..10M}, with values comparable to the \Kepler\ orbital cyclic frequency $\alyas\,\simeq\,31.1\,$nHz, due to its revolution period ($\simeq$ 372.57 days).
This frequency can be spuriously introduced as an alias of genuine frequencies,  when the photometric time series presents recurrent gaps introduced when one or more \Kepler\ quarters are regularly missing. This occurs  when a star falls on a damaged CCD \citep{2014MNRAS.445.2698H}.  Therefore, it is necessary to take care to avoid stars with poor duty cycle.

The diversity of the origin of the stars previously analyzed prevents any firm information about potential bias selection when using them. For instance, \cite{2014A&A...572L...5M} favored the analysis of stars with large masses in order to get a more detailed view on the secondary red clump. Therefore, in order to reduce the selection bias as much as possible, stars identified as RC stars were systematically analyzed.

The data set we obtained is shown in the $\Dnu$ -- $\Tg$ diagram, where the large separation, $\Dnu,$ is representative of the envelope density and the period spacing, $\Tg$, of the core property (Fig.~\ref{fig-dnuDPi}). This diagram allows us to infer the evolutionary stage of evolved low-mass stars.

\section{Mixed mode rotational splittings\label{analysis}}

\subsection{Global fit}

The global fit of the mixed-mode pattern relies first on the identification of radial modes \citep{2009A&A...508..877M} in order to precisely locate  pure pressure dipole modes that would still exist in the absence of coupling. This location must be precise enough to include the variation with frequency of the large separation due to second-order effects and to the modulation of pressure modes by acoustic glitches \citep{2015A&A...579A..84V}. Then, the parameters of the gravity component of the mixed modes can be measured: the period spacing, $\Tg$, and the coupling factor, $q,$ are automatically determined, whereas the gravity offset, $\epsg$, is derived from the identification of a given mixed mode. The final step of the global fit comes from the identification of the rotational splittings
\citep{2018A&A...618A.109M}.

\subsection{Rotational splittings}

Both mean core and mean envelope rotation rates can be derived from dipole mixed modes \citep{2013A&A...549A..75G}. Here, we used previous measurements to check that the mean envelope rotation is significantly smaller than the mean core rotation, aside from a few secondary clump stars \citep{2015A&A...580A..96D}. Measuring the envelope rotation is however difficult and noisy.  For sake of simplicity, in the following, we assume that the  mean envelope rotation can be neglected in first approximation.

\begin{figure}
\begin{center}
\includegraphics[width=5.8cm]{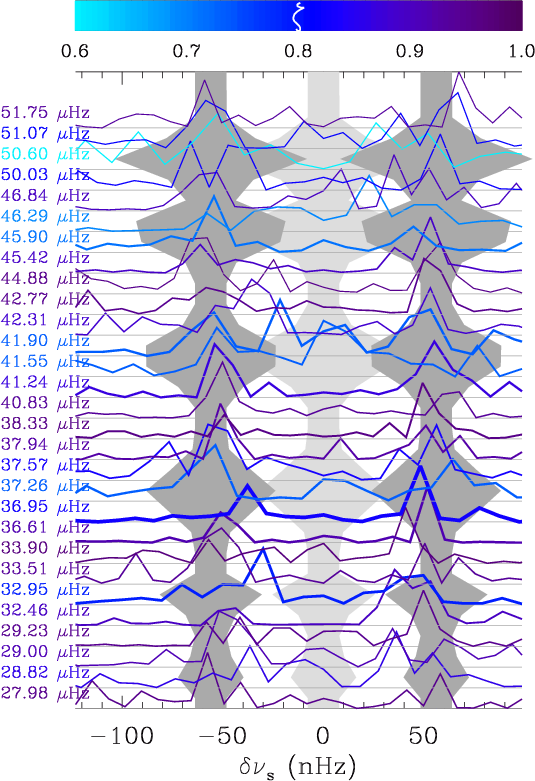}
\caption{Power density spectrum of rotational $\ell=1$ multiplets in a typical red clump star (KIC 4459025). Each multiplet is centered on its mean frequency, indicated on the left side; the y-axis shows the different multiplets, with normalized heights, ranked in frequency. The thickness of the lines is scaled by the mean height of the multiplets. The color codes the mean parameter $\zeta$ of the multiplet: gravity-dominated (pressure-dominated) mixed modes appear in purple (blue), respectively. The grey areas indicate the expected width of the components of the multiplets \citep{2018A&A...618A.109M}.}
\label{fig-echelle}
\end{center}
\end{figure}

A convenient way to get rid of the complexity of the mixed-mode frequency pattern consists of considering stretched spectra, constructed in such a way that mixed modes are expected to be regularly spaced. Here, contrary to the method used for inferring regular period spacings \citep{2015A&A...584A..50M}, this approached is aimed at unveiling regular frequency spacings and thus making use of stretched frequencies, $\nus$, defined by: \begin{equation}\label{eqt-stretchfreq}
  \diff \nus = {\diff \nu \over \zeta}
  ,
\end{equation}
where $\zeta$ is the stretching function. Then, stretched rotational multiplets exhibit the same equidistance $\dnurot$ , which can be measured by the autocorrelation of the stretched spectrum. A typical identification is shown in Fig.~\ref{fig-echelle}, where the identification is complicated by the stochastic excitation of the peak; however, the value of the rotational splitting is secured by the many multiplets that are observed.

For most stars in the red clump (and contrary to RGB stars), rotational splittings are small enough to avoid any confusion with period spacings due to the overlap of different modes with different mixed-mode orders. However, two effects, combined with the stochastic excitation of modes, can complicate the identification of rotational doublets or triplets. First, for dim magnitudes or crowded fields, oscillation spectra show a poor signal. Second, low values of the inclination angle hamper the observation of $|m|=1$ modes. A combination of the two effects complicates the identification further.

\begin{figure}
\includegraphics[width=8.8cm]{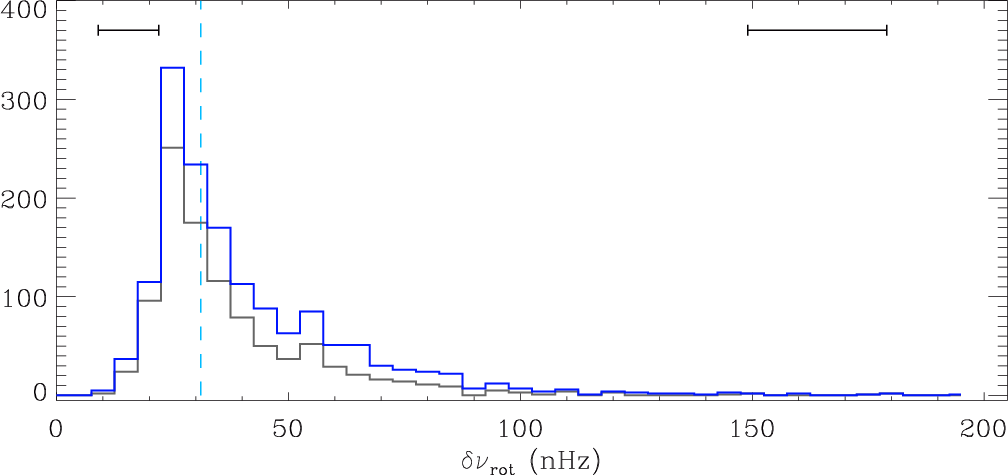}
\caption{Histograms of $\dnurot$ in red-clump stars (blue curve for all stars; grey curve for those systematically analyzed). The vertical dashed line at 31.1 nHz shows the \Kepler\ orbital frequency. Typical uncertainties are provided by the horizontal bars, for small and large splittings.}\label{fig-histonurot}
\end{figure}

We could derive $\Ntotal$ rotational splittings in RC stars, among which more than 850 were analysed in a systematic way, enlarging the previous catalog by a factor larger than six. We checked whether the rotational signatures close to the \Kepler\ year alias $\alyas$ could be artifacts, which is not the case since many signatures near $\alyas$ are due to doublets separated by $2 \alyas$, for stars seen nearly edge-on. We also specifically checked cases where missing \Kepler\ observation quarters could introduce an alias signature. We could then infer the absence of spurious signatures due to the \Kepler\ orbit, as shown by the histogram of the rotational splittings  (Fig.~\ref{fig-histonurot}), where the peak at low values is due to the most numerous low-mass RC stars, and is not affected by the \Kepler\ orbital frequency.

\begin{figure}
\includegraphics[width=8.9cm]{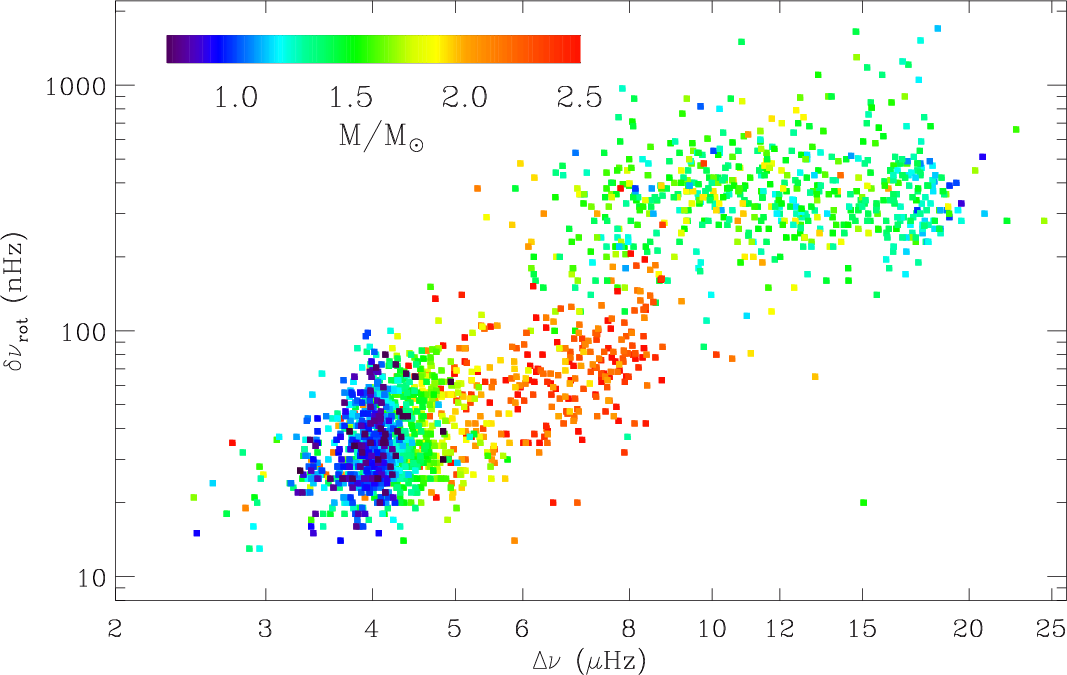}
\caption{Rotational splittings $\dnurot$ as a function of the large separation $\Dnu$. Large symbols represent red-clump stars, whereas small symbols are stars on the RGB. The color codes the stellar mass. Typical uncertainties for clump stars vary from 12\,nHz at low $\Dnu$  to 25\,nHz at large $\Dnu$.}\label{fig-Dnunurot}
\end{figure}

\subsection{Biases and completeness}


The method we used for measuring the rotational splittings is efficient up to a value corresponding to twice the spacing between consecutive mixed modes, as tested for RGB stars. For RC stars, this corresponds to values much above the distribution shown by the histogram and proves the completeness of this distribution at large values of $\dnurot$.
The situation at the low end of the distribution of the rotational splittings is less clear, with rotational splittings down to the minimum measured value $\dnurot = 12$\,nHz, slightly above the frequency resolution of 7.8\,nHz reached by the four-year time series. Lower splittings cannot be measured. Furthermore, low values of simple splittings in rotational triplets for low-inclinations stars are also missing, contrary to double splittings of rotational doublets of stars seen edge-on. These effects explain the decrease of the distribution of $\dnurot$ in the histogram for low splittings. However, the missing information is limited: the number of stars with a clear oscillation signal and not showing any rotational splitting is low, that is, below 15\,\%. This indicates that a vast majority of rotational splittings could be measured, so that the histogram in Fig.~\ref{fig-histonurot} is representative of the core rotation rates, but it fails at giving clear indication of the smallest core rotation rates.

\subsection{Core rotation rates along stellar evolution}

In order to study the core rotation along stellar evolution and its dependence on stellar parameters, we considered the rotational splittings as a function of the large separation $\Dnu$ (Fig.~\ref{fig-Dnunurot}). We use the seismic scaling relations to derive proxies of the stellar mass, with the additional information of effective temperatures \citep{2017AJ....154...94M}.

Core rotations of RC stars are clearly distinct from the values measured for stars on the RGB (Fig.~\ref{fig-Dnunurot}). Contrary to the flat distribution on the RGB shown by \cite{2018A&A...616A..24G}, the fit for red-clump stars shows a clear dependence on the large separation,
\begin{equation}\label{eqt-scaling-Dnu}
  \dnurot \simeq (5.21 \pm0.13)\ \Dnu^{1.322\pm0.039}
  ,
\end{equation}
for $\Dnu$ in $\mu$Hz and $\dnurot$ in nHz.

\begin{figure}
\includegraphics[width=8.9cm]{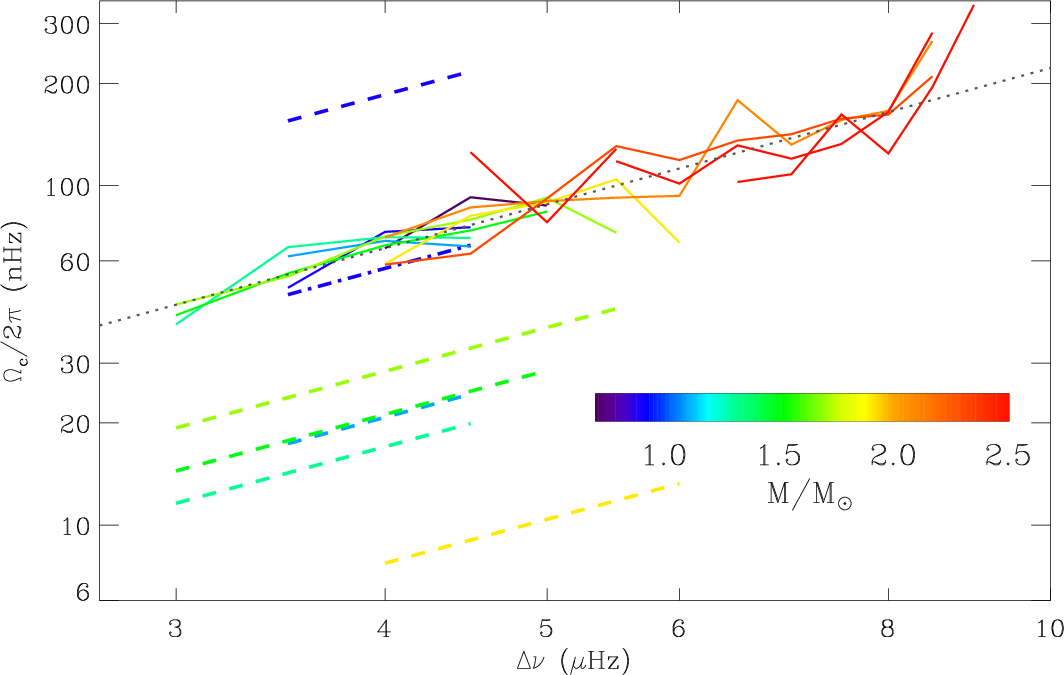}
\caption{Mean evolution of the core rotational splittings of RC stars as a function of the stellar mass, with bins of 0.2 \,$\Msun$ (full lines, with the same color code as in Fig.~\ref{fig-Dnunurot}).
The black dotted line shows a scaling as $\Dnu^{4/3}$.
Extrapolations of the mean rotation from the MS rotation rates are shown for similar mass bins (dashed lines). The blue dot-dashed line, for the $[0.7, 0.9] \Msun$ mass bin, takes a 0.2-$\Msun$ mass loss into account. }\label{fig-evolMSRC}
\end{figure}

In order to emphasize any mass dependence, we investigated the variation of median values of the rotational splittings in different bins of mass, as a function of $\Dnu$ (Fig.~\ref{fig-evolMSRC}). All curves are close to each other and close to the mean power law in $\Dnu,$ expressed by Eq.~(\ref{eqt-scaling-Dnu}). The variations of the parameters (exponent, factor) with stellar masses are not significant. Seismic parameters and rotational splittings are listed in Table \ref{tab-param}. The full table is available online.

\section{Discussion\label{discussion}}

In this section, ensemble asteroseismology is considered to derive information about the transfer of angular momentum from the measurements of the core rotation rate. Then, the comparison of our measurement with the surface rotation rates of MS stars is used to provide a consistent picture of the mean stellar rotation along stellar evolution.

\begin{table*}
\caption{Seismic parameters}\label{tab-param}
\begin{tabular}{rrrrrr}
\hline
 KIC     & $\numax\qquad$      & $\Dnu\qquad$       & $\Tg\qquad$ & $q\qquad$   & $\dnurot\quad$ \\
         & ($\mu$Hz$)\qquad$   & ($\mu$Hz$)\qquad$  & (s$)\qquad$ &             & (nHz$)\quad$   \\
  \hline
 1026084 &$ 45.77\pm 0.58$&$ 4.45\pm0.03$&$ 283.3\pm 1.7$&$0.35\pm0.04$&$  41\pm 11$\\
 1161618 &$ 34.06\pm 0.53$&$ 4.08\pm0.03$&$ 329.5\pm 1.7$&$0.19\pm0.03$&$  45\pm 11$\\
 1162746 &$ 27.20\pm 0.50$&$ 3.81\pm0.03$&$ 279.3\pm 2.4$&$0.30\pm0.04$&$  65\pm 18$\\
 1163453 &$ 39.79\pm 0.58$&$ 4.43\pm0.03$&$ 301.0\pm 1.6$&$0.26\pm0.03$&$  25\pm  9$\\
 1163621 &$ 51.06\pm 0.65$&$ 5.00\pm0.04$&$ 331.5\pm 1.1$&$0.25\pm0.03$&$  47\pm 10$\\
 1297272 &$ 31.34\pm 0.55$&$ 4.19\pm0.03$&$ 321.0\pm 3.7$&$0.38\pm0.05$&$  43\pm 16$\\
 1430912 &$ 32.66\pm 0.53$&$ 4.04\pm0.03$&$ 319.0\pm 1.9$&$0.25\pm0.03$&$  25\pm 10$\\
 1431316 &$ 37.20\pm 0.56$&$ 4.31\pm0.03$&$ 322.1\pm 1.3$&$0.33\pm0.03$&$  36\pm 10$\\
 1435573 &$ 25.38\pm 0.47$&$ 3.59\pm0.03$&$ 290.0\pm 2.0$&$0.28\pm0.04$&$  23\pm 10$\\
  \hline
\end{tabular}

List of the full data set with $\Ntotal$ red-clump stars is available on line as a
CDS/VizieR document.
\end{table*}

\subsection{Rotation and stellar evolution\label{subsec-evol}}

The power law (Eq.~\ref{eqt-scaling-Dnu}) used to fit the mean core rotation rate is compatible with an evolution as $\dnurot \propto \Dnu^{4/3}$. According to seismic scaling relations \citep[$\Dnu\propto M^{1/2}\,R^{-3/2}$, e.g.][]{2011A&A...530A.142B,2012MNRAS.419.2077M}, this corresponds to the core rotation period  scaling as
\begin{equation}\label{eqt-scaling}
  T\ind{core} \simeq T_0   \ \left({M\over \Msun}\right)^{-2/3} \left({R\over R_\odot}\right)^{2}
  ,
\end{equation}
where $M$ and $R$ are, respectively, the stellar mass and radius; and $T_0 \simeq 1.62$ days. From our modelling, we found that there is no scaling relation between $T\ind{core}$ and the core radius (Appendix \ref{app-model}). Thus, the observations indicate that the mean core rotation period evolves as $R^{2}$. A similar strong correlation was found in secondary-clump stars \citep[Fig. 6 of][]{2019ApJ...887..203T}.

\subsection{Locking of the core and envelope rotation rates\label{subsec-lockdown}}

In fact, such a relationship between the rotation and the stellar radius also exists for the mean envelope rotation, as derived from a simple analytical three-zone model composed of a rigid, dense helium core, an intermediate radiative layer with a rotation profile allowed to vary smoothly, and a rigid convective envelope. This analytical view is presented in Appendix \ref{app-analytical}. For such a model, the total angular momentum writes:
\begin{equation}\label{eqt-J}
 \momcin = \fac\, \Mc\, R^2 \Omega\ind{env}
  ,
\end{equation}
where $\Mc = 4 \pi \rho\ind{c} \rHBS^3 /3$ is the core mass below the helium-burning shell of radius, $\rHBS$, and $\fac$ is a structural parameter that accounts for the non-uniform mass distribution inside the helium core. This parameter $\fac$ is expected not to vary significantly during the red clump evolution. Therefore, supposing the conservation of $\momcin$  and the predominant sensitivity of $\momcin$ to the radius variations yields
\begin{equation}\label{eqt-rotenv}
   \Omega\ind{env} \propto \Mc^{-1} R^{-2}
   .
\end{equation}
With both core and envelope rotation rates evolving together as $R^{-2}$, we conclude that differential rotation is locked in core-helium burning red giants:
\begin{equation}\label{eqt-lock}
   \Omega\ind{env} \propto \Omega\ind{core}
   .
\end{equation}
The variation of the mean core rotation period as $R^{2}$ indicates that a mechanism for transferring angular momentum from the core to surface is able to couple the stellar radius and the core rotation. Such a transfer mechanism exists on the RGB, but is not efficient enough for spinning down their core rotation while the envelope radius drastically increases
\citep{2018A&A...616A..24G}.

Evolution in the RC is much slower than on the RGB, by a factor of 80 for a 1.0-$\Msun$ star or a factor of 15 for 1.8-$\Msun$ star, according to the modelling. So, the measurement of the core rotation of RC stars enables us to constrain the efficiency of the mechanism(s) transferring angular momentum over a longer evolutionary phase: for RGB stars, the mechanism must ensure a stable core rotation period; for RC stars, it must ensure a rotation rate locked to the stellar radius, assuming the conservation of the global angular momentum.

\subsection{Extrapolation from the main sequence\label{extrapol}}

Since the core rotation is tightly coupled with the surface rotation, we also investigated the possible relationship between the core rotation in RC stars and the envelope rotation in MS stars \citep{2014ApJS..210....1C,2014A&A...572A..34G}, in the limit case of solid rotation. The significant mass redistribution inside the star along stellar evolution was taken into account with MESA simulations, which we used to infer the variations of the ratio $\coefI$ used for expressing moments of inertia as a function of the stellar fundamental parameters (Appendix \ref{sec-moments}):
\begin{equation}\label{eqt-mom}
  \momin = \coefI \ M\, R^2
  .
\end{equation}
As solid rotation is assumed, the envelope rotation $\Omega\ind{MS}$ is representative of the mean rotation. From the conservation of the total angular momentum, and assuming negligible mass loss, we can extrapolate the mean rotation $\bar\Omega$ during the clump phase from $\Omega\ind{MS}$:
\begin{equation}\label{eqt-moment}
  \bar\Omega = \Omega\ind{MS}\, {\coefI\ind{MS} \over \coefI\ind{RC}}
  \left({R\ind{MS} \over \Rsun} \right)^2
  \left({M \over \Msun} \right)^{-2/3}
  \left({\Dnu \over \Dnu_\odot} \right)^{4/3}
  ,
\end{equation}
where the radius, $R\ind{MS}$, and the rotation rate, $\Omega\ind{MS}$, are derived from the catalogs of MS rotation, while the coefficients $\coefI\ind{MS}$ and $ \coefI\ind{RC}$ are derived from MESA.
For stars above 1.1\,$\Msun$, the mean rotation rates extrapolated from the MS are much slower than the core rotation measured in the red clump (Fig.~\ref{fig-evolMSRC}). However, this is not the case in the range of $[0.7-0.9]\,\Msun$, where extrapolated mean rotation rates are much significantly larger. In order to explain this apparently paradoxical situation for the lowest mass,
two solutions are possible, which both involve mass loss: a strong mass loss may evacuate a large amount of angular momentum; the too high inferred rotation may come from an overestimated mass redistribution due to a too low mass in the MS. Assuming a 0.2 $\Msun$ mass loss in this second scenario, the change in the coefficient $\coefI\ind{RC}$, which expresses the mass redistribution in Eq.~(\ref{eqt-moment}), moves the extrapolated value (Fig.~\ref{fig-evolMSRC}) toward a trend that agrees with other evolutions. These simple arguments point toward a mass loss of at least 0.2\,$\Msun$ for the less massive stars of our data set. Extensive modelling to confirm this result is beyond the scope of this work.

\section{Conclusion\label{conclusion}}

We measured rotational splittings for $\Ntotal$ core-helium burning stars observed by \Kepler. The evolution of their mean core rotation rates scales approximately as $\Dnu^{4/3}$, which corresponds to an evolution in $R^{-2}$ dominated by the expansion of the stellar envelope. This indicates that mechanisms for transferring angular momentum from the core to the envelope are efficient in the core-helium burning phase.

Assuming the global conservation
of  angular momentum, a simple analytical model allows us to infer that the envelope rotation rate also evolves as $R^{-2}$ in core-helium burning stars.
Therefore, we infer that the rotation rates in the core and in the envelope jointly evolve as $R^{-2}$. Despite this efficient spinning down of the core associated with the expansion of the envelope, stars in the red clump do not rotate rigidly, but with a core rotation and an envelope rotation that are tightly coupled. The extrapolation of the surface rotation of MS stars provides similar information. For stars with a mass below 0.9\,$\Msun$, a mass loss of about 0.2\,$\Msun$ at the tip of the RGB appears necessary to reconcile the extrapolated values and compare them with values in the helium-burning phase.

\begin{acknowledgements}
We thank the entire \emph{Kepler} team, whose efforts made these results possible. During this work, C. P. was supported by a postdoctoral grant from Centre National d'Etudes Spatiales (CNES). G. D. gratefully acknowledges support from the Australian Research Council through Discovery Project DP190100666. S. D. acknowledges support from the project BEAMING ANR-18-CE31-0001 of the French National Research Agency (ANR) and from the Centre National d’Etudes Spatiales (CNES).
\end{acknowledgements}

\bibliographystyle{aa} 
\bibliography{48338}

\begin{appendix}

\section{Modelling\label{app-model}}

Our study highlights the dependence of the core rotation on the stellar radius. We also used our modelling to test the possible role of the radiative-core radius. Therefore, we constructed stellar models with the evolution code MESA \citep[][and references therein]{2019ApJS..243...10P}, which fit the seismic properties of red-clump stars. To do so, the models included He-core overshooting, following the prescription of \cite{2017MNRAS.469.4718B} and motivated by the initial study of \cite{2015MNRAS.453.2290B}. Their exact description is given in \cite{2022A&A...668A.115D}.
\begin{figure}
\includegraphics[width=8.8cm]{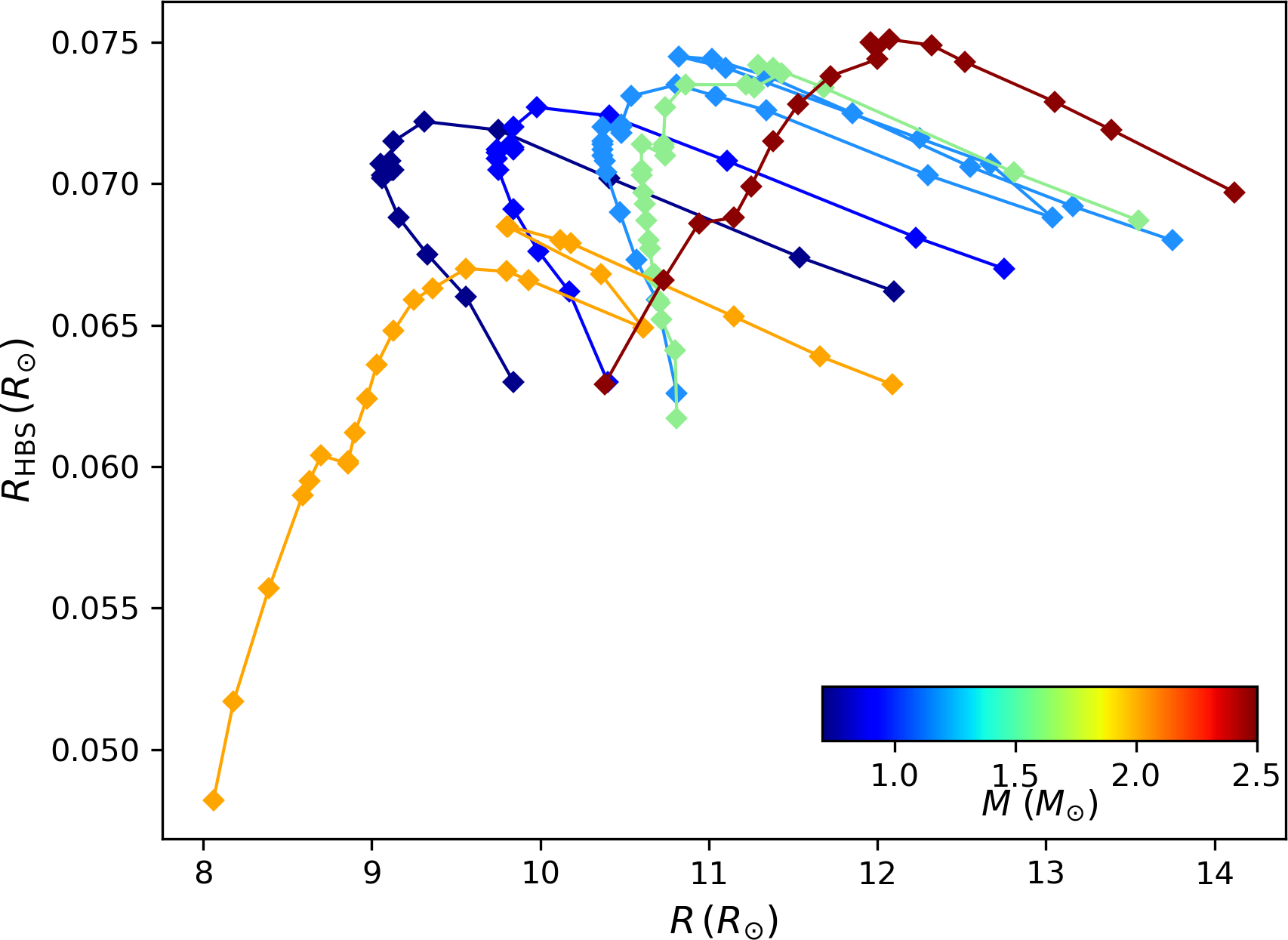}
\caption{Radius of the radiative core, estimated as the radius of the helium-burning shell, as a function of the stellar radius, for various stellar masses.}\label{fig-RRC}
\end{figure}
From this modelling, we can derive that the location of the hydrogen-burning shell of stars with a degenerate core ($M\leq 1.8\,\Msun$) does not significantly vary when stars evolve in the red clump and displays a non-monotonic behavior (Fig.~\ref{fig-RRC}). This underlines the fact that the evolution of the core radius is dominated by the variation of the stellar radius.

\section{Analytical rotation model\label{app-analytical}}

\subsection{Two-zone rotation model\label{app-2D}}

For an evolved star with a dense core and a large envelope, respectively below and above the hydrogen-burning shell (at radius $\rHBS$), the density can be approximated as:
\begin{equation}\label{eqt-densd}
   \rho =
    \left\{
    \begin{array}{cl}
      \rho\ind{c}   & \mbox{for } r < \rHBS , \\
      \rho\ind{c}\, (r / \rHBS)^{-3} & \mbox{for } r \ge\rHBS .\\
    \end{array}
   \right.
\end{equation}
Assuming a shellular rotation profile $\Omega(r)$, the total angular momentum is:
\begin{eqnarray}
  \momcin &=&  \int_0^R \rho r^2 \sin^2 \theta \; \Omega(r) \; r^2 \sin \theta \, \diff \theta \diff \varphi \diff r \\
    &=& \frac{8 \pi }{3} \int_0^R \rho r^4 \Omega(r) \, \diff r \label{eqt-profilJ}
    .
\end{eqnarray}
Such a power law for the density is justified to first approximation in a large part of the region above the helium core where the core-to-envelope density contrast is high, as demonstrated by simple homologous analyses \citep[e.g.][]{2012sse..book.....K,2020A&A...634A..68P}. This approximation certainly fails in the near-surface layers where the density drastically drops, but the contribution of this thin region to the following computations is expected to be negligible, and thus should not modify the general result.

If we suppose a uniform rotation profile in both regions,
\begin{equation}\label{eqt-omegadef}
  \Omega =
   \left\{
   \begin{array}{ll}
      \Omega\ind{c}   & \mbox{for } r < \rHBS \\
      \Omega\ind{env} & \mbox{for } r \ge\rHBS \\
    \end{array}
  \right.
  ,
\end{equation}
the core contribution to $\momcin$ appears negligible, due to the fact that $\rHBS \ll R$. So, from Eqs.~(\ref{eqt-densd}) and (\ref{eqt-profilJ})
we get:
\begin{equation}\label{eqt-J-Omegaenv}
   \momcin  \simeq \Mc\, R^2 \Omega\ind{env}
  ,
\end{equation}
where $\Mc = 4 \pi \rho\ind{c} \rHBS^3 /3$ is the core mass below $\rHBS$.
The conservation of $\momcin$ then gives
\begin{equation}\label{eqt-rotenv}
   \Omega\ind{env} \propto \Mc^{-1} R^{-2}
   .
\end{equation}
We can then retrieve a dependence, as introduced by Eq.~(\ref{eqt-J}). In fact, a constant density in the helium core as supposed in Eq.~(\ref{eqt-densd}) is not rigorously met in red clump stars where the density already starts decreasing between the convective inner layers and the hydrogen-burning shell, in a significant way. As a result, the total angular momentum is lower than predicted by Eq.~(\ref{eqt-J-Omegaenv}) and can be rewritten as $ \momcin  = \fac \Mc R^2 \Omega\ind{env}$ instead. The  structural parameter $\fac$ accounts for the drop in density inside the helium core and is expected to be nearly constant as a function of $M_{\rm c}$. We note that, except for this structural parameter, $\momcin$ keeps its dependence in $R^2$ since the contribution of the outer envelope to the total angular momentum remains predominant, owing to the large core-envelope density contrast.

\subsection{Three-zone rotation model\label{app-3D}}

The hypotheses in the two-zone model can be relaxed in order to account for different rotation profiles in the upper radiative zone and in the convective envelope. We therefore considered a modified rotation profile, with a transition region between the hydrogen-burning shell and the base of the convective region (at radius $\rBCZ$), expressed by
\begin{equation}\label{eq-hbs_bcz}
  \Omega = \Omegaenv\left({\rBCZ \over r}\right)^{\expoO}
  \hbox{ \ for \ } \rHBS < r < \rBCZ .
\end{equation}
The total angular moment is written as
\begin{eqnarray}
  \momcin &\simeq&  {2\over 5} \Mc \rHBS^2 \Omegac
  + 2 \Mc  {\rBCZ^\expoO \over \rHBS^{\expoO -2}} \Omegaenv \int_1^{\rBCZ / \rHBS} x^{1-\expoO} \diff x \nonumber\\
   &+& 2 \Mc \rHBS^2 \Omegaenv \int_{\rBCZ / \rHBS}^{R / \rHBS} x \diff x
   .
\end{eqnarray}
Then,  for $\expoO \ne 2$ (final conclusion is unchanged in the case $\expoO = 2$, owing to the properties of the logarithm),
\begin{eqnarray}
  \momcin &\simeq&  2 \Mc \rHBS^2
  \left[ {\Omegac \over 5} + {\Omegaenv\over 2-\expoO}
  \left(\left( {\rBCZ\over \rHBS} \right)^{2} -  \left( {\rBCZ\over \rHBS} \right)^{\expoO}\right)
  \right.\nonumber\\
   &+& \left. {\Omegaenv \over 2}
   \left(\left( {R\over \rHBS} \right)^{2} - \left( {\rBCZ\over \rHBS} \right)^{2}\right)
   \right]
    .
\end{eqnarray}
This shows that the two-zone model is retrieved when $\expoO=0$. From our modelling, we note that $\rHBS \ll \rBCZ \ll R$, so that
\begin{equation}\label{eq-3Dc}
   \momcin \simeq  \Mc
   \Omegaenv  R^{2}
   .
\end{equation}
Hence, the total angular momentum mainly depends on the core mass and on the envelope rotation (Eq.~(\ref{eqt-J-Omegaenv})).

\section{Mass redistribution\label{sec-moments}}

\begin{figure}
\includegraphics[width=8.8cm]{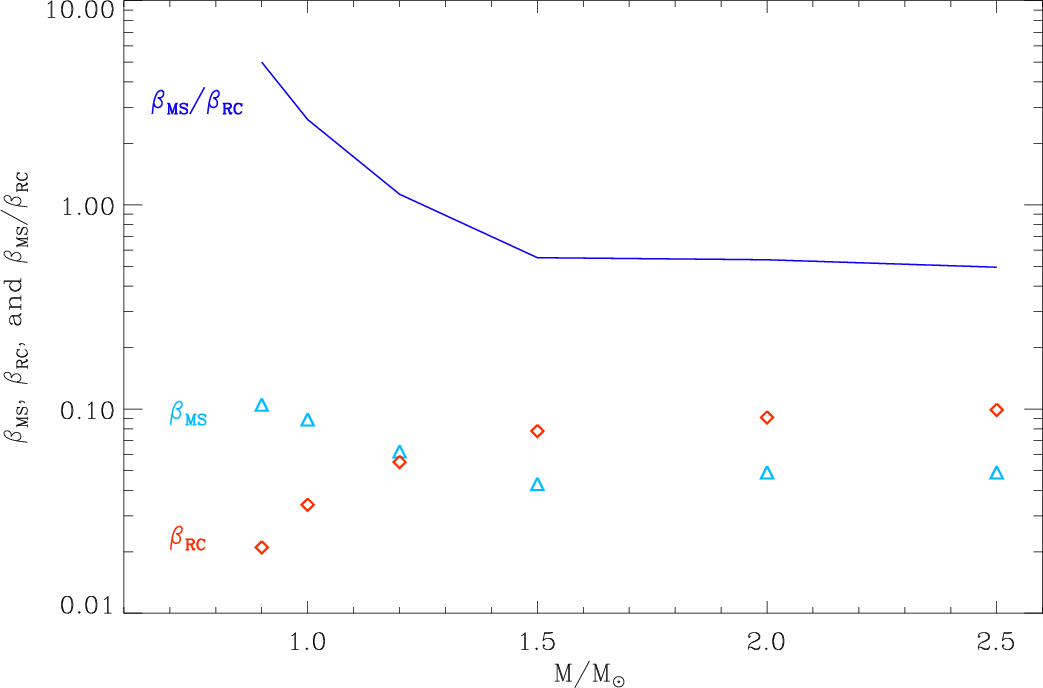}
\caption{Dimensionless coefficients $\coefI$ of the moment of inertia, in units $M R^2$, on the MS ($\coefI\ind{MS}$, light blue triangles) and in the red clump ($\coefI\ind{RC}$, red diamonds), derived from our modelling. The blue curve shows the variation of their ratio $\coefI\ind{MS} / \coefI\ind{RC}$.}
\label{fig-mom}
\end{figure}

The modelling presented in Appendix \ref{app-model} was also used to infer the coefficients, $\coefI,$ that express the mass redistribution in the stars according to Eq.~(\ref{eqt-mom}). The values for MS stars are taken at the ZAMS; for clump stars, they correspond to a central helium mass fraction of 0.5; typical variations in both evolutionary stages are 35\,\%. From these coefficients, we derived the ratio $\coefI\ind{MS} / \coefI\ind{RC}$ introduced by Eq.~(\ref{eqt-moment}), shown in Fig.~\ref{fig-mom}. Large values of this ratio, for the less massive stars, indicate the important core shrinking from MS to RC; this shrinking is an order of magnitude less important for more massive stars.

\end{appendix}
\end{document}